\documentclass[12pt]{article}

\usepackage{tabulary,graphicx,amsmath,amsfonts,amssymb}
\usepackage[utf8x]{inputenc}
\usepackage[T1]{fontenc}
\usepackage{siunitx}
\usepackage{cite}
\usepackage{amsmath}
\usepackage{subfigure}
\usepackage{caption}
\usepackage{graphicx}
\usepackage{bm}
\usepackage{float}
\usepackage{url,multirow,morefloats,floatflt,cancel,tfrupee}
\makeatletter
\usepackage{multicol}
\usepackage{colortbl}
\usepackage{xcolor}
\usepackage{pifont}
\usepackage[nointegrals]{wasysym}
\usepackage{abstract}
 
\title{\textbf{\Large Dual-comb spectroscopy over 100 km open-air path}}
\date{}

\begin{document}

\maketitle
\setlength{\parindent}{0pt}
\author
{
	Jin-Jian Han\textsuperscript{1,2,3,*}, Wei Zhong\textsuperscript{4,*}, Ruo-Can Zhao\textsuperscript{1,3,4,5,*}, 
	Ting Zeng\textsuperscript{1,2,3}, Min Li\textsuperscript{1,2,3}, 
	Jian Lu\textsuperscript{1,2,3}, Xin-Xin Peng\textsuperscript{1,2,3}, Xi-Ping Shi\textsuperscript{6}, Qin Yin\textsuperscript{4}, Yong Wang\textsuperscript{7},
	Ali Esamdin\textsuperscript{7}, Qi Shen\textsuperscript{1,2,3}, Jian-Yu Guan\textsuperscript{1,2,3}, Lei Hou\textsuperscript{1,2,3}, Ji-Gang Ren\textsuperscript{1,2,3},
	Jian-Jun Jia\textsuperscript{3,8}, Yu Wang\textsuperscript{3,4,5}, Hai-Feng Jiang\textsuperscript{1,2,3}, Xiang-Hui Xue\textsuperscript{1,3,4,5}, Qiang Zhang\textsuperscript{1,2,3,8},
	Xian-Kang Dou\textsuperscript{1,3,4,5}, Jian-Wei Pan\textsuperscript{1,2,3}\\
	\\
	\normalsize{$^{1}$Hefei National Research Center for Physical Sciences at the Microscale,University of Science and Technology of China, Hefei, China, }\\
	\\
	\normalsize{$^{2}$Shanghai Research Center for Quantum Science and CAS Center for Excellence in Quantum
		Information and Quantum Physics,University of Science and Technology of China, Shanghai 201315, China}\\
		\\
    \normalsize{$^{3}$Hefei National Laboratory, University of Science and Technology of China, Hefei, China}\\
    \\
    \normalsize{$^{4}$CAS Key Laboratory of Geospace Environment, School of Earth and Space Sciences, University of Science and Technology of China, Hefei, China}\\
     \\
    \normalsize{$^{5}$Anhui Mengcheng Geophysics National Observation and Research Station, University of Science and Technology of China}\\
    \\
    \normalsize{$^{6}$Faculty of Information Science and Engineering, Ningbo University, Ningbo, China}\\
    \\
     \normalsize{$^{7}$Xinjiang Astronomical Observatory, Chinese Academy of Sciences, Urumqi, China}\\
     \\
     \normalsize{$^\ast$These authors contributed equally to this work.}
}
\begin{abstract}

Satellite-based greenhouse gases (GHG) sensing technologies play a critical role in the study of global carbon emissions and climate change. However, none of the existing satellite-based GHG sensing technologies can achieve the measurement of broad bandwidth, high temporal-spatial resolution, and high sensitivity at the same time. Recently, dual-comb spectroscopy (DCS) has been proposed as a superior candidate technology for GHG sensing because it can measure broadband spectra with high temporal-spatial resolution and high sensitivity. The main barrier to DCS’s display on satellites is its short measurement distance in open air achieved thus far. Prior research has not been able to implement DCS over 20 km of open-air path. Here, by developing a bistatic setup using time-frequency dissemination and high-power optical frequency combs, we have implemented DCS over a 113 km turbulent horizontal open-air path. Our experiment successfully measured GHG with 7 nm spectral bandwidth and a 10 kHz frequency and achieved a CO2 sensing precision of <2 ppm in 5 minutes and <0.6 ppm in 36 minutes. Our results represent a significant step towards advancing the implementation of DCS as a satellite-based technology and improving technologies for GHG monitoring. 
\end{abstract}
\section{Introduction}
\begin{multicols}{2}
	Atmospheric spectroscopy is a key technique used to study the chemical and physical properties of the Earth's atmosphere by analyzing the interaction of light with atmospheric molecules and particles. This technique, especially when it’s satellite-based, has become a valuable tool in the study of atmospheric composition, structure, and dynamics, which has essential applications in global climate change, carbon budget assessment, deep space exploration, and air pollution research.

	To date, atmospheric spectroscopy techniques such as grating spectrometers\cite{chatterjeeInfluenceNinoAtmospheric2017,elderingOrbitingCarbonObservatory22017,liuContrastingCarbonCycle2017}, Integrated Path Differential Absorption Lidar (IPDA Lidar)\cite{ehretSpaceborneRemoteSensing2008}, heterodyne spectroradiometers\cite{rodinHighResolutionHeterodyne2014,rodinVerticalWindProfiling2020}, and Fourier transform spectrometers (FTS)\cite{hammerlingGlobalCO2Distributions2012} have been able to provide remote ground-based or satellite-based gas spectroscopy\cite{bauwensImpactCoronavirusOutbreak2020,boersmaNearrealTimeRetrieval2007,butzAccurateCO2CH42011,clerbauxMonitoringAtmosphericComposition2009,hedeliusSouthernCaliforniaMegacity2018,wangCarbonDioxideRetrieval2020,wuSectorbasedAttributionUsing2022,zhangCarbondioxideAbsorptionSpectroscopy2021} with various temporal and spatial resolutions and sensitivities.

	Among these methods, grating spectrometers, FTS, and heterodyne spectroradiometers require sunlight and can’t work at night. And they suffer from scarce measurements at high latitudes. IPDA lidar, typically employing a narrow-linewidth laser, has difficulty measuring multiple gas species. Tunable laser remote spectroscopy suffers from long measurement time and atmospheric turbulence. To realize global coverage, current satellite-based instruments are usually in low orbit (typically like A-Train), resulting in a long repeat cycle of 16 days\cite{behrenfeldAnnualBoomBust2017,behrenfeldGlobalSatelliteobservedDaily2019,crispOnorbitPerformanceOrbiting2017}. New methods with fast and continuous measurements are needed to fulfill the study of point carbon sources and sinks in high temporal-spatial resolution.  Recently, open-air DCS\cite{coburnRegionalTracegasSource2018a,cossel28KmOpen2021,giorgettaBroadbandPhaseSpectroscopy2015a,hermanPreciseMultispeciesAgricultural2021a,riekerFrequencycombbasedRemoteSensing2014a,waxmanEstimatingVehicleCarbon2019a,waxmanIntercomparisonOpenpathTrace2017a,ycasMidinfraredDualcombSpectroscopy2019a,yunSpatiallyResolvedMass2022a}, as a novel coherent spectroscopic tool, has proven to be a superior candidate technology for precise, accurate, and continuous fast multi-gas measurements. DCS can cover a broadband of 1-10 µm, allowing the realization of active remote sensing of large numbers of gas species using only a single dual comb system.

	Since the first demonstrations, numerous DCS experiments have been spawned.  The technology has been exploited in oil field monitoring\cite{coburnRegionalTracegasSource2018a}, urban vehicular GHG emission\cite{waxmanEstimatingVehicleCarbon2019a}, livestock emission measurement\cite{hermanPreciseMultispeciesAgricultural2021a}, urban GHG monitoring\cite{cossel28KmOpen2021}, etc. Owing to high acquisition speed for a whole interferogram, multiple comb mode with absolute accuracy, stability with negligible linewidth, and high-brightness coherent light sources, DCS has demonstrated its advantages of immunity to turbulent speckle and background noise, resulting in the capability of sensing longer distances without calibration, which is considered to be a perfect precision spectroscopy tool for remote sensing of the atmosphere. Furtherly, by developing networks consisting of satellite-ground links and horizontal links, the spatial distribution and variations of GHG could be acquired, contributing significantly to global climate change research\cite{mitchellLongtermUrbanCarbon2018a,yoshidaRetrievalAlgorithmCO22011}. By establishing of a DCS linking ground stations and Geosynchronous Earth orbit (GEO) satellites, it will be possible to monitor point carbon sources and sinks continuously and precisely\cite{schwandnerSpaceborneDetectionLocalized2017}.

	However, the longest reported distance for DCS is limited to less than 20 km\cite{cossel28KmOpen2021}, with a link loss of 40-60 dB. This is lower than the typical channel loss of 60-80 dB for high-orbit satellite-to-ground links using single-mode fiber coupling\cite{caldwellQuantumlimitedOpticalTime2023,shenExperimentalSimulationTime2021b,renGroundtosatelliteQuantumTeleportation2017}. Meanwhile, as will be discussed later, the existing configuration of the DCS experiment with folded path measurement proves impractical for the satellite-to-ground link due to the significant geometric loss exceeding 100 dB. Thus, the demonstration of DCS over long distances with high channel loss in the horizontal open air becomes inevitable and urgently required for satellite-based applications\cite{madhusudhanExoplanetaryAtmospheresKey2019}. In addition, DCS measurements over a 100 km horizontal range can cover any city, allowing long-term monitoring of changes in water vapor and carbon dioxide levels. This will provide insight into the impact of urban land surfaces and the effects of urbanization on the environment\cite{johnsonEvolutionLifeUrban2017,luoThermodynamicAnalysisAirground2019}.
\end{multicols}
\section{Setup}
\begin{multicols}{2}
	The main obstacle to longer distance of DCS is the low signal-to-noise ratio. We try to increase the signal-to-noise ratio on both the protocol and hardware sides. Fig. 1A shows the current monostatic DCS protoco\cite{coddingtonDualcombSpectroscopy2016a}. Two optical combs are locked to the same terminal by sharing a common frequency standard, and a large mirror is placed on the other remote terminal as a reflector. Since the frequency lines in the two combs are locked together, any absorption or dispersion will bring influence to the interference of the two combs. DCS utilizes the change of interference to figure out the atmospheric property. In the past, there were two ways to achieve interference: One is that the dual combs interfere before passing through the atmosphere. The other is that one comb first passes through the atmosphere first and then interferes with the local comb. The former can only provide the amplitude spectrum, while the latter can provide both amplitude and phase spectrum. Nonetheless, both ways pass through the measured atmosphere path twice, causing attenuation and turbulence twice. Moreover, the laser beam expands with distance and the mirror is usually not large enough to reflect all of the beam, causing additional geometric loss. For example, after passing through 100 km of turbulent atmosphere, it is normal for the beam to spread to a radius of 3 m, and mirror of this size are expensive and impractical. This substantially limits their measurement range. Indeed, this issue turns out to be insoluble for satellite-ground links. Assuming a telescope aperture of 1 m and a mirror aperture of 0.5 m—a logical assumption considering the manufacturing process and satellite placement—the anticipated geometric attenuation reaches a substantial 105.8 dB for a high orbit spanning a satellite-ground distance of 36000 km (GEO) (see Supplementary Material for details).

	To increase the signal-to-noise ratio, we exploit a bistatic protocol, as shown in Fig. 1B. The two combs are separated into two remote terminals. Both combs are sent to each other through the atmosphere and interfere with each other at the receiving terminals. With this configuration, both absorption and phase spectrum change by the atmosphere over the path can be achieved at each terminal. Since the comb passes through the atmosphere path for only once and no reflector is required, the link loss can be significantly reduced compared to the previous monostatic round-trip protocol.

	However, this new protocol introduces a new problem. The basis of DCS is that the two combs need to share a common frequency standard. In the monostatic protocol, this condition is easy to achieve. Usually, a local frequency standard, such as an Rb atomic clock, is exploited to lock both combs. In the bistatic protocol, each comb has its own frequency standard and the frequency difference and fluctuation of the two frequency standards will deteriorate the interference and reduce the frequency accuracy of the spectrum. Compared to the monostatic protocol, the frequency accuracy of the bistatic protocol is worse by a factor $M=f_{r}/\Delta f_{r}$  , where $f_{r}$ is the repetition rate of the local comb and $\Delta f_{r}$ is the difference between the repetition rate of the two combs (see Supplementary Material for details). Fig. 1C shows an example of spectra with two frequency standards, where a frequency offset has occurred. Note that the offset is different for each measurement since the state of the frequency standard can change in the long run. To maintain the same frequency accuracy as the monostatic protocol, the frequency standard must be remotely locked. This problem can be naturally solved by the time-frequency dissemination technology\cite{bauchComparisonFrequencyStandards2006b,caldwellQuantumlimitedOpticalTime2023,drosteOpticalFrequencyTransferSingleSpan2013b,giorgettaOpticalTwowayTime2013b,predehl920KilometerOpticalFiber2012,shenExperimentalSimulationTime2021b,shenFreespaceDisseminationTime2022b}. Fig. 1C also shows that the frequency offset is eliminated when the frequency standard is traced back to one terminal by frequency comparison.
\end{multicols}

	\begin{figure}[H]
		\centering
		\includegraphics[scale=1]{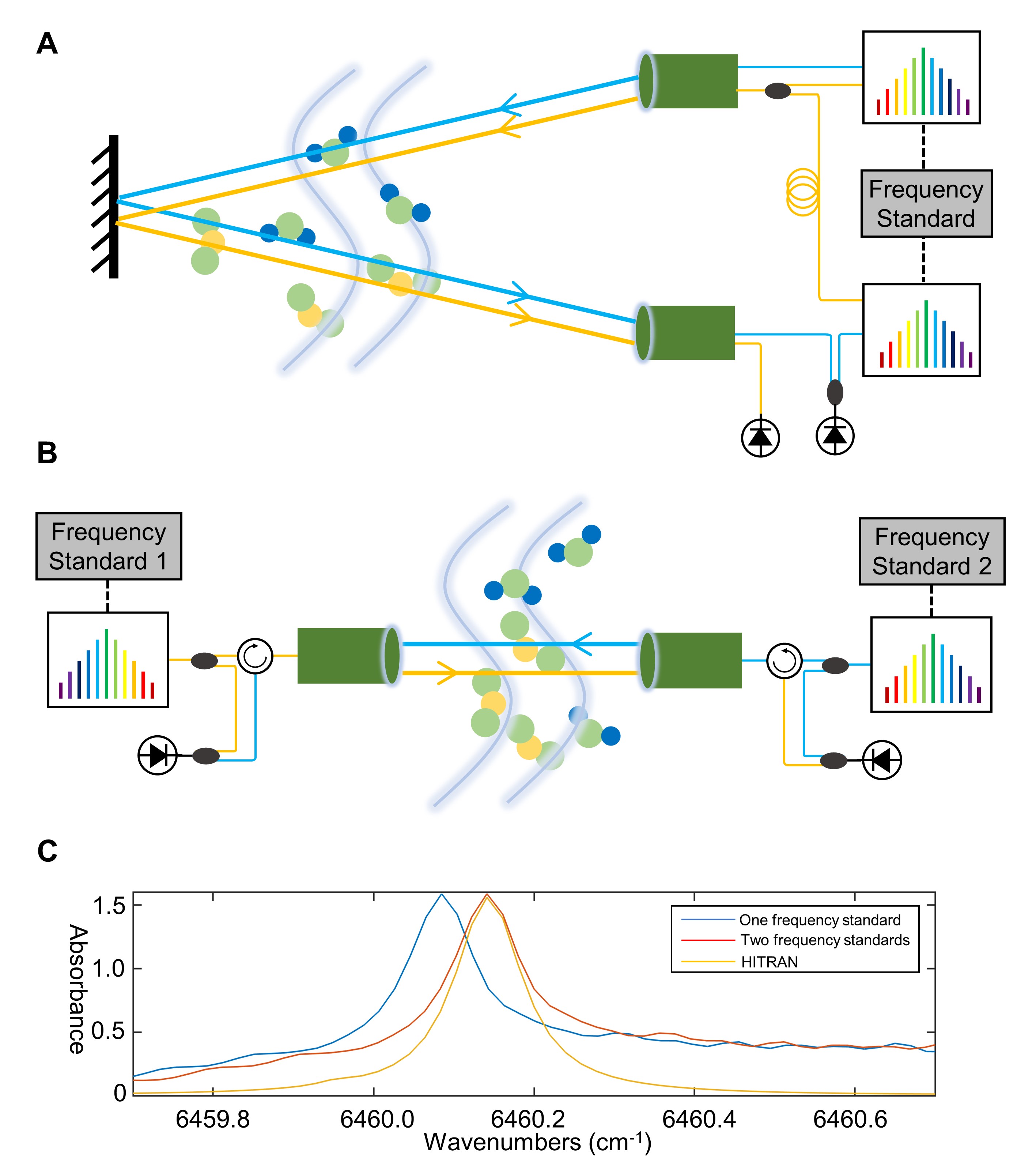}
		\caption{{\bfseries{Protocols of DCS.}} (A) Two classic configurations used in monostatic DCS protocol: Dual combs interfere before passing through the atmosphere (orange lines) and one comb first passes through the atmosphere and then interferes with the local comb (blue lines). (B) Our bistatic two-way protocol: Two combs separated into two remote terminals are sent to each other through the atmosphere and interfere with each other at the receiving terminals.  (C) Comparison of measured spectra with HITRAN. Two frequency standards cause a frequency offset of about 1.1 GHz (bule line).  The frequency standard has been traced back to either terminal using time-frequency transfer, and the frequency offset was eliminated (red line).}
	\end{figure}

\begin{multicols}{2}
	There are many ways to achieve frequency comparison. In our experiment, both frequency comparison and spectral information are obtained from the interferograms without the need for a separate link. A total of 1024 points around the interferogram centerburst are used for Fourier transform to extract the smoothed phase spectrum, which is used for phase compensation and frequency comparison. When the comb from terminal B is transmitted through the atmosphere to terminal A, there is a delay $\tau_{AB}$ . Let $\angle F(\nu,0)$ and $\angle F(\nu,\tau_{AB})$  represent the spectral phase after Fourier transforms of the first interferogram and other interferograms, respectively.  Here  $\tau$ is the frequency obtained from the Fourier transform of the interferogram. The delay  $\tau_{AB}$ is proportional to the difference between  $\angle F(\nu,0)$  and $\angle F(\nu,\tau_{AB})$ , i.e.,
	\begin{equation}
		\begin{aligned}
			\Delta \angle F(\nu,\tau_{AB})&=\angle F(\nu,0)-\angle F(\nu,\tau_{AB}) \\
			&=k \nu \cdot \tau_{AB}+b
		\end{aligned}
	\end{equation}
	Where $k$ and $b$ are constants (see Supplementary Material for details). Therefore,  $\tau_{AB}$ can be calculated by fitting a line to $\Delta \angle F(\nu,\tau_{AB})$ . Note that this delay is affected by both the drift between the frequency standards and by the atmosphere, so it varies with time. The two-way method is usually used to counter the influence of the atmosphere:
	\begin{equation}
		\begin{aligned}
			\tau_{AB}&=\tau_{link}-\tau  \\
			\tau_{BA}&=\tau_{link}+\tau
		\end{aligned}
	\end{equation}
	
	So,
	\begin{equation}
		\begin{aligned}
			\tau=\frac{1}{2} (\tau_{BA}-\tau_{AB})
		\end{aligned}
	\end{equation}
	
	Where  $\tau_{link}$ is the delay of atmosphere and $\tau$ is the time fluctuations of frequency standards. The fractional frequency deviation is the derivative of $\tau$ and frequency comparison can be calculated in real time.
	The signal-to-noise ratio of a signal interferogram is too low to observe the spectral information. From Eq. (1), we can get,
	\begin{equation}
		\angle F(\nu,0)=\angle F(\nu,\tau_{AB})+\Delta \angle F(\nu,\tau_{AB})
	\end{equation}
	
	A linear fitting of $\nu$ represents $\Delta \angle F(\nu,\tau_{AB})$. All interferograms used for calculation have the same delay as the first interferogram after adding $\Delta \angle F(\nu,\tau_{AB})$, and phase compensation is achieved.

	Figure 2 illustrates our experimental setup. Two terminals are located at Nanshan and Gaoyazi, Urumqi, Xinjiang Province, which are more than 113 km apart.

	Each terminal has a rubidium clock (SRS FS725, 10 MHz) that is used as the absolute frequency standard. Each comb is phase-locked to the local ultrastable laser (USL, Stable Laser Systems SLS-INT-1550-1) with sub 1 Hz linewidth at the optical frequency of 193.4 THz, while the carrier-envelope offset frequency $f_{CEO}$  is locked by way of the f-2f method\cite{baumann20YearsDevelopments2019}. Both frequency instabilities are reached $1\times 10^{-21}@10000$s . The repetition rates  $f_{r}$ of the combs, used as the reference signals of the local sampling and reference clock of all electronics, are 250 MHz and 250 MHz+2.5 kHz at Nanshan and Gaoyazi, respectively.
\end{multicols}
\begin{figure}[H]
	\centering
	\includegraphics[scale=0.8]{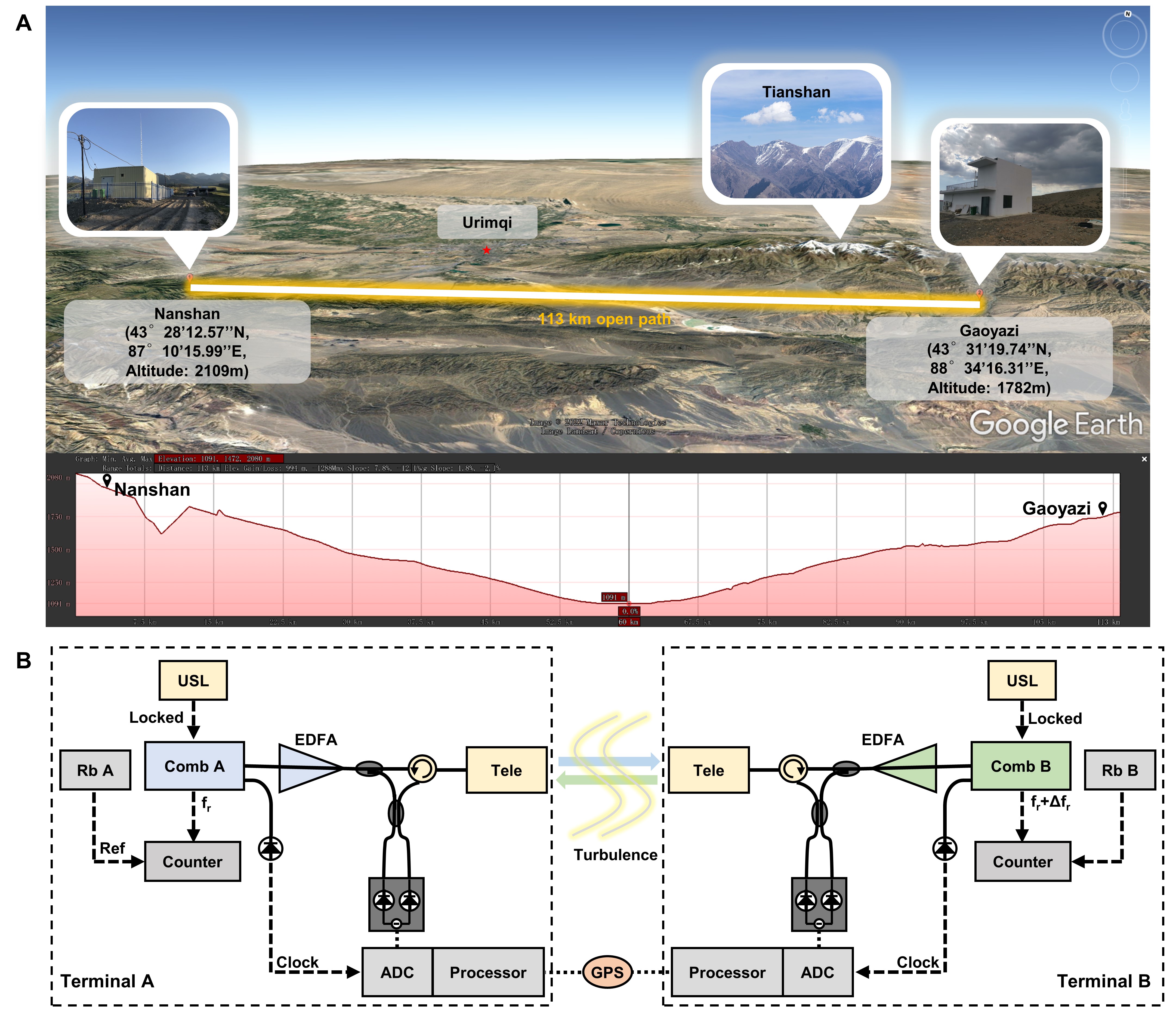}
	\caption{{\bfseries{The experimental setup.}} (A) Overview of the 113 km DCS. (B) The main configuration and setup at each terminal. Rb: Rubidium clock; USL: Ultrastable laser; EDFA: Erbium-doped fiber amplifier; Tele: Telescope; ADC:  Analog-to-Digital Converter; GPS: Global positioning system}
\end{figure}
\begin{multicols}{2}
	We develop a high-power optical comb and ultrasensitive detection setup to cope with the large fading of the link. The output power of each comb can reach 1 W after amplification in a two-stage high-power erbium-doped fiber amplifier (EDFA) with a 20-nm filtering spectrum bandwidth. The 3-dB bandwidth is approximately 7 nm centered at 1,545 nm. To avoid damage caused by high power density, a chirped pulse with 60 ps-100 ps is produced. Water chillers are used to avoid damage to the EDFAs due to overheating. Highly stable and efficient optical transmission systems, including polarization-maintaining fibers and dedicated optical telescopes, push the detected power into the nanowatt range.

	The optical paths before the telescope are fiber paths using polarization-maintaining fibers, most of which are integrated into an aluminum box. To maintain its stability, we controlled the temperature using a thermoelectric cooler (TEC), and the standard deviation of the temperature was 7 mK. The Global Positioning System (GPS) is used to synchronize the data collection start time, which is within 30 ns. Both improve the accuracy of the time-frequency transfer\cite{shenFreespaceDisseminationTime2022b}.
	The power of the received signal swings widely due to the characteristics of free space links over 100 km, so it is essential to determine when the signal is received and to observe the signal-to-noise ratio of the interferograms in real time. Here, a threshold-triggered acquisition method is introduced: Once the detector voltage exceeds a fixed threshold, an interferogram of 41,350 points centered around the centerburst is captured. This method reduces many post-processing processes and, more importantly, reduces the pressure on memory and transmission bandwidth.

	Two homemade optical telescopes with an aperture of 400 mm and a focal length of 1600 mm are used to transmit and receive the laser. A direction-tracking system with a beacon laser and an adjustable mirror is developed to overcome large link loss at each telescope.
\end{multicols}

\section{Results}
\begin{multicols}{2}
	In our experiment, to achieve the same frequency accuracy as the monostatic protocol, the accuracy of the frequency comparison must be less than 0.1 Hz, and our previous work has confirmed that we can achieve this\cite{shenFreespaceDisseminationTime2022b}. The frequency accuracy can be significantly improved by more than 5 orders, which is 1 GHz before and 10 kHz after the frequency comparison. To ensure that most of the phase unwrapping is correct, the return signal of the first interferogram is above 100 nW and is above 5 nW for other interferograms, equivalent to a total attenuation of 83 dB.

	As mentioned above, a window of 41,350 points centered around the interferogram centerburst is captured, so the point spacing is approximately 605 MHz rather than 250 MHz for $\frac{f_{r}}{\Delta f_{r}}=100,000$ points. The spectra span 6,000 comb teeth covering 6505.5 to 6455.5 cm$^{-1}$. Fig. 3A and Fig. 3B show an example of the normalized intensity spectrum and the phase spectrum after subtraction of quadratic polynomials attributed to the air dispersion, acquired over 1 hour, allowing long-term coherent averaging.

	Molecular absorbance and phase signature scale linearly with concentration so that we can extract the gas concentration directly from the absorbance and the phase spectra separately. The reference spectrum cannot be obtained because the two combs cannot be placed on one terminal. A penalty least squares algorithm\cite{zhangBaselineCorrectionUsing2010a} (see Supplementary Material for details) is used to convert the absorbance or phase spectra to concentration. The model used for fitting is the High-Resolution Transmission Molecular Absorption Database (HITRAN) 2020, the complex Voigt line shape\cite{gordonHITRAN2020MolecularSpectroscopic2022}. The path-averaged atmospheric pressure and temperature for the inversion are the averages of the sensors at both terminals. Fig. 3C and Fig. 3D show the resulting absorption spectrum and phase spectra.
\end{multicols}
\begin{figure}[H]
	\centering
	\includegraphics[scale=1]{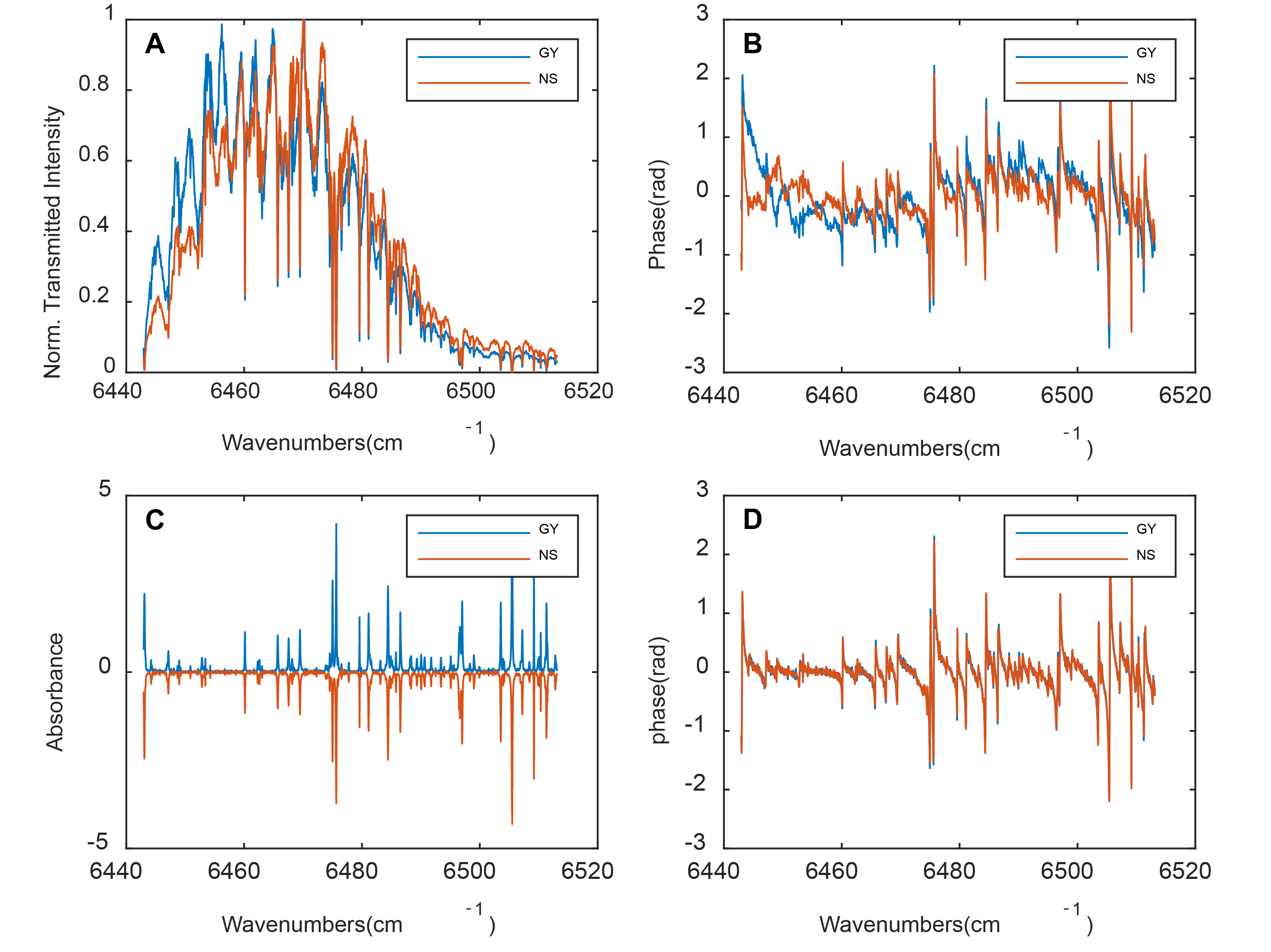}
	\caption{{\bfseries{Original spectrum and resulting spectrum.}} (A) The normalized intensity spectrum. (B) The phase spectrum after subtracting quadratic polynomials. (C) The resulting absorbance. (D) The resulting phase spectrum.}
\end{figure}
\begin{multicols}{2}	
	
	Figure 4 shows the CO$_{2}$ concentration and H$_{2}$O concentrations extracted independently from the absorbance and phase spectra at both terminals at 5-minute intervals from June 5$^{th}$ to 7$^{th}$. The changing trends of the four retrievals complied well. A slow increase in CO$_{2}$ concentration from evening to early morning was observed in the path we measured with little human activity, which was attributed to plant respiration. The water vapor concentration was high on June 5$^{th}$ after the rainy day on June 4$^{th}$ and decreased significantly after the sunny day on June 6$^{th}$. The CO$_{2}$ concentration from both absorbances agrees withins $\pm$10 ppm, 2 ppm on average, while the H$_{2}$O concentration agrees within $\pm$300 ppm, 36 ppm on average. The sensitivity of the CO$_{2}$ concentration is calculated throughout the relative stability from 2022-06-06 23:29 to 2022-06-07 04:29, as shown in Fig. 5. All of them are above 2 ppm in 5 minutes and above 0.6 ppm in 36 minutes.
\end{multicols}	
\begin{figure}[H]
	\centering
	\includegraphics[scale=1]{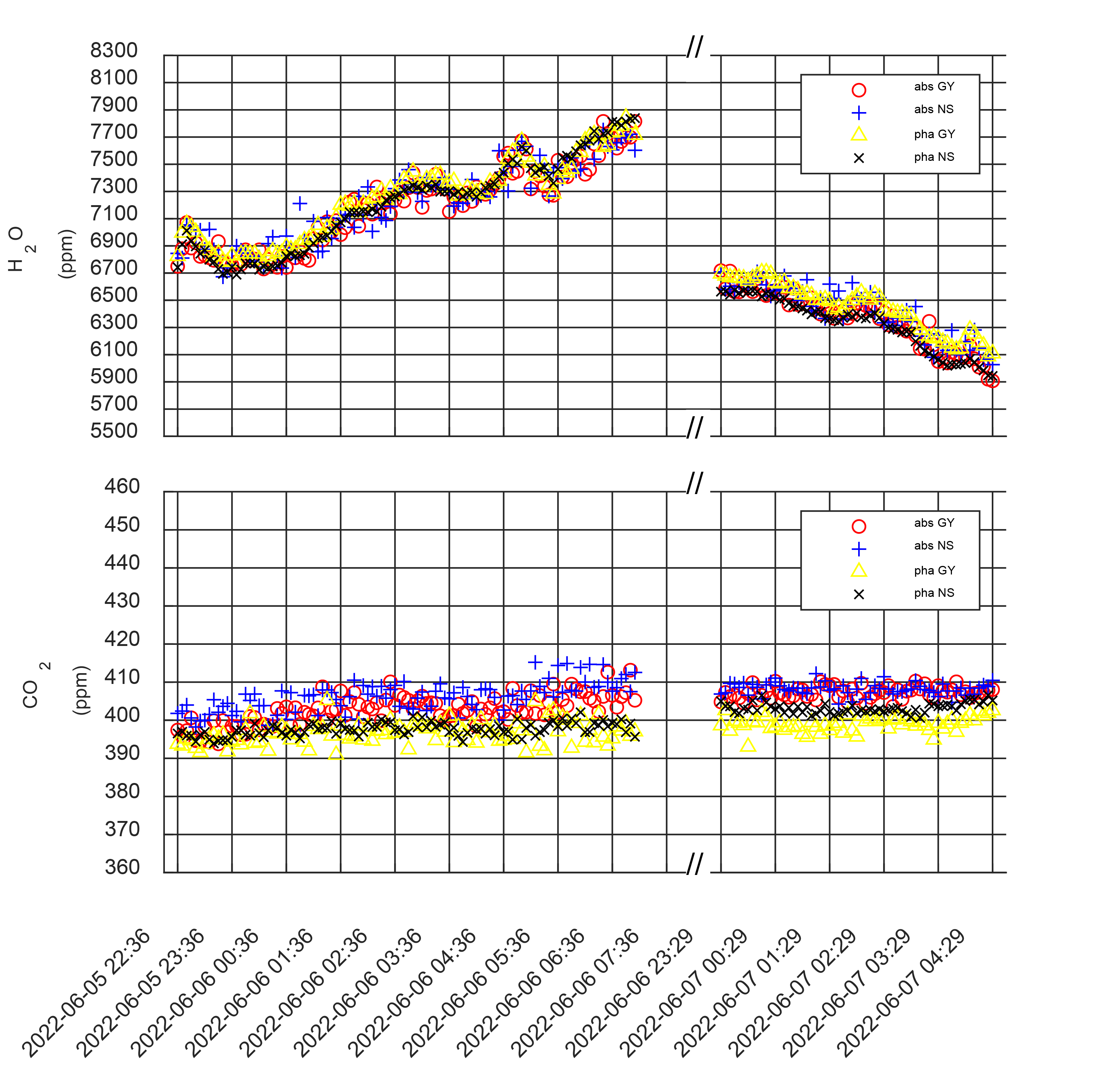}
	\caption{{\bfseries{CO$_2$ concentration and H$_2$O concentration extracted independently from the absorbance and phase spectra at both terminals.}} The changing trends of the four retrievals complied well.
	}
\end{figure}
\begin{multicols}{2}	
	The offset of the concentration extracted from the absorbance and phase spectra for both terminals occurs. We attribute this to two reasons. First, coarse atmospheric parameters introduce different offsets to the inversion of the intensity and phase spectra. Second, scattering in the optical path is unavoidable at high output power. The scattered light and the signal light will also interfere, influencing the inversion results (see Supplementary Material for details). Future work on retrieving atmospheric parameters with higher accuracy from atmospheric models or absorbance\cite{riekerFrequencycombbasedRemoteSensing2014a} and the cepstral-domain fitting technique\cite{coleBaselinefreeQuantitativeAbsorption2019a} will make the inversion more accurate.

\end{multicols}
	\begin{figure}[H]
	\centering
	\includegraphics[scale=1]{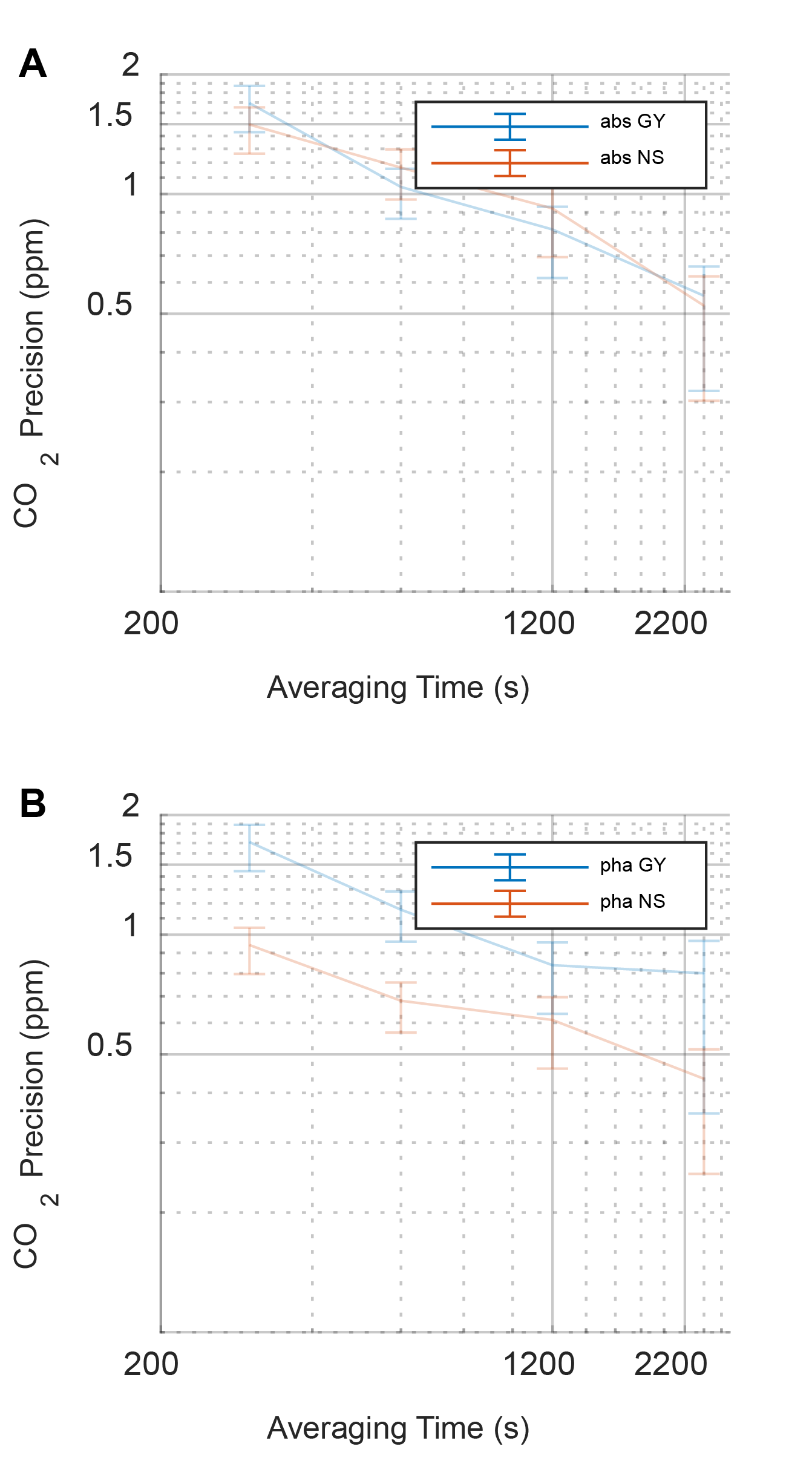}
	\caption{{\bfseries{Allan deviation for the four retrievals over a period of relative stability.}} (A) Allan deviation of concentration retrievals from absorbance. (B) Allan deviation of concentration retrievals from the phase.}
\end{figure}
\section{Conclusion}
\begin{multicols}{2}
	In summary, we demonstrate a bistatic dual-comb spectroscopy protocol combining time-frequency transfer technology. Several essential techniques have been employed and verified, particularly time-frequency techniques, high-power stable optical frequency combs, and high-efficiency transceiver telescopes. This protocol is more advantageous for long-range measurements. The full complex spectrum has been measured at both terminals over a 113 km open-air path, with a channel loss of 83 dB. The retrieved concentration precision of CO2 is \textless 2 ppm in 5 minutes and \textless0.6 ppm in 36 minutes. This paves the way for satellite-ground dual-comb spectroscopy measurements with high resolution, broad spectrum, and high sensitivity. Based on our work, we anticipate that an ultra-long-range network of interlinked ground-space-based open-path DCS systems will play an important role in predicting global climate change and supporting the development of accurate absorption models by continuously monitoring global GHG fluxes over multiple spatial and temporal scales.
\end{multicols}
\section*{Acknowledgements}
The author would like to thank Shui-Ming Hu and Yan Tan from University of Science and Technology of China. 
This research was supported by the National Key Research and Development Programme of China (grant no. 2020YFA0309800, 2020YFC2200103); Strategic Priority Research Programme of Chinese Academy of Sciences (grant no. XDB35030000); National Natural Science Foundation of China (grant no. T2125010, 61825505, 42188101, 42125402); Anhui Initiative in Quantum Information Technologies (grant no. AHY010100); Key R$\&$D Plan of Shandong Province (grant no. 2020CXGC010105, 2021ZDPT01); Shanghai Municipal Science and Technology Major Project (grant 2019SHZDZX01); Innovation Programme for Quantum Science and Technology (grant no. 2021ZD0300100, 2021ZD0300300); Fundamental Research Funds for the Central Universities.

\clearpage
\section*{Supplementary Materials and methods}
\setcounter{equation}{0}
\renewcommand{\theequation}{S.\arabic{equation}}
\renewcommand{\thetable}{S.\arabic{table}}
\setcounter{figure}{0}
\renewcommand{\thefigure}{S.\arabic{figure}}
\textbf{The loss for satellite-ground link.} 
The link loss can be calculated using
\begin{equation}
	\eta ={{\eta }_{tele\_s}}{{\eta }_{tele\_g}}{{\left( \frac{{{D}_{rec}}}{L\theta } \right)}^{2}}{{T}_{atm}}{{\eta }_{fiber}}
\end{equation}

Table S1 explains the meaning and empirical values of each symbol. The parameter values are based on \cite{shenExperimentalSimulationTime2021b}. Using the above values, the uplink loss is calculated to be 77.1 dB while the downlink loss is 54.9 dB for an effective transmitter full-angle divergence   of 5 urad and a fiber coupling efficiency of 0.15.
\begin{table}[h]
	\centering
	\begin{tabular}{c|c|c}
		\hline
		\textbf{Symbols} & \textbf{Parameters} & \textbf{Values} \\
		\hline
		${{\eta }_{tele\_s}}$ & The telescope optical efficiency of the satellite & 0.8 \\
		${{\eta }_{tele\_g}}$ & The telescope optical efficiency of the ground & 0.8 \\
		$\theta $ & The effective transmitter full-angle divergence & 15 urad (uplink) \\
		$L$ & The satellite-to-ground distance & 36000 km \\
		${{D}_{rec}}$ & The aperture of the receive telescope & 0.5 m \\
		${{T}_{atm}}$ & The atmospheric transmittance & 0.7 \\
		${{\eta }_{fiber}}$ & The fiber coupling efficiency & 0.05 (uplink) \\
		\hline
	\end{tabular}
	\caption{Meaning of each parameter in Equation S.1 and empirical values}
\end{table}

\textbf{The geometrical loss for a high orbit satellite-ground folding link.} 
Here we calculate the geometrical loss for a high orbit satellite-ground folding link where a reflector is placed on the ground or on a satellite (GEO).
The geometrical efficiency can be expressed as: 
\begin{equation}
	{{\eta }^{geo}}={{\left( \frac{D}{L\theta } \right)}^{2}}
\end{equation}
where $D$ is the aperture of telescope or reflector, $L$ is the satellite-to-ground distance, $\theta$  represent the effective transmitter full-angle divergence for the downlink or the uplink. For folding link, the total geometrical loss is the product of uplink loss and downlink loss 
\begin{equation}
	\eta _{total}^{geo}=\eta _{up}^{geo}\cdot \eta _{down}^{geo}={{\left( \frac{{{D}_{mirr\_s}}}{L\cdot {{\theta }_{up}}} \right)}^{2}}\cdot {{\left( \frac{{{D}_{tele\_g}}}{L\cdot {{\theta }_{down}}} \right)}^{2}}                    
\end{equation}

The values of parameters could be estimated as Table S2, and the loss is calculated to be 105.8 dB.

\begin{table}[H]
	\centering
	\begin{tabular}{c|c|c}
		\hline
		\textbf{Symbols} & \textbf{Parameters} & \textbf{Values} \\
		\hline
		$L$ & The satellite-to-ground distance & 36000 km \\
		${{\theta }_{up}}$ & The effective transmitter full-angle divergence for the uplink & 15 urad \\
		${{\theta }_{down}}$ & The effective transmitter full-angle divergence for the downlink & 5 urad \\
		${{D}_{tele\text{ }\!\!\_\!\!\text{ g}}}$ & The aperture of telescope on ground & 1m \\
		${{D}_{mirr\text{ }\!\!\_\!\!\text{ s}}}$ & The aperture of telescope on satellite & 0.5 m \\
		\hline
	\end{tabular}
	\caption{Meaning of each parameter in Equation S.3 and empirical values}
\end{table}

\textbf{The frequency accuracy of one frequency standard versus two frequency standards.} When all phase-locked loops are closed for each comb, we have the relation ${{f}_{CEO}}+N\cdot {{f}_{r}}+{{f}_{Beat}}={{f}_{CW}}$, where ${{f}_{CW}}$ is the CW frequency of the ultrastable laser (USL), ${{f}_{r}}$ is the repetition of the comb, and $N$ is the index of the tooth of comb locking to USL. In our experiment, the carrier-envelope offset frequency ${{f}_{CEO}}$ and the beat frequency ${{f}_{Beat}}$ are set to be equal but in opposite directions for each terminal so that ${{f}_{CEO}}+{{f}_{Beat}}=0$, thus $N\cdot {{f}_{r}}={{f}_{CW}}$. Figure S3 shows the locking direction of them at both terminals. Using the local rubidium clock as a reference, the frequency of the USL at each terminal is known by counting the repetition frequency of the comb.

Within the Nyquist sampling bandwidth, the beat frequency between two adjacent comb teeth is:
\begin{equation}
	{{\nu }_{RF}}=2{{f}_{Beat}}-n\cdot \Delta {{f}_{r}}-\Delta {{f}_{CW}} 
\end{equation}

$\Delta {{f}_{r}}={{f}_{r,GY}}-{{f}_{r,NS}}$, $\Delta {{f}_{CW}}={{f}_{CW,GY}}-{{f}_{CW,NS}}$. $n$ is an integer representing the position of the comb teeth.
The complex spectrum corresponding to the intensity and phase of pairs of comb teeth reflects the atmosphere's response to the signal comb. So, the beat frequency, i.e. radiofrequency, is needed to be converted to optical frequency. The optical frequencies corresponding to the nth pair of comb teeth are, respectively:
\begin{equation}
	{{\nu }_{OF,GY}}={{f}_{CW,GY}}-{{f}_{Beat}}+n\cdot {{f}_{r,GY}} 
\end{equation}
\begin{equation}
	{{\nu }_{OF,NS}}={{f}_{CW,NS}}+{{f}_{Beat}}+n\cdot {{f}_{r,NS}} 
\end{equation}
From equation (S.4), equation (S.5), and equation (S.6), we get:
\begin{equation}
	{{\nu }_{OF,GY}}=(2{{f}_{Beat}}-{{\nu }_{RF}}-\Delta {{f}_{CW}})\cdot \frac{{{f}_{r,GY}}}{\Delta {{f}_{r}}}+{{f}_{CW,GY}}-{{f}_{Beat}}
\end{equation}
\begin{equation}
	{{\nu }_{OF,NS}}=(2{{f}_{Beat}}-{{\nu }_{RF}}-\Delta {{f}_{CW}})\cdot \frac{{{f}_{r,NS}}}{\Delta {{f}_{r}}}+{{f}_{CW,NS}}+{{f}_{Beat}}
\end{equation}
The absolute frequency of the USL at each terminal is calculated based on the local rubidium clock, and the frequency accuracy is 10 kHz. Therefore, the frequency accuracy of $\Delta {{f}_{CW}}$ is the same order of magnitude. However, a factor of $M=\frac{{{f}_{r}}}{\Delta {{f}_{r}}}$ multiplied by $\Delta {{f}_{CW}}$, which is 100,000 in our experiment, significantly deteriorates the frequency accuracy. 

For monostatic DCS protocol, where only one laser is used to lock two combs, $\Delta {{f}_{CW}}$ is zero.

\begin{table}[h]
	\centering
	\begin{tabular}{c|c|c}
		\hline
		\textbf{Symbols} & \textbf{Parameters} & \textbf{Values} \\
		\hline
		$f_{CEO,NS}$ & Carrier-envelope offset frequency of the comb at Nanshan & 35MHz \\
		$f_{Beat,NS}$ & The beat frequency of the comb at Nanshan & -35MHz \\
		$f_{r,NS}$ & The repetition rate of the comb at Nanshan & ~250MHz \\
		$N_{NS}$ & Index of the comb's tooth locking to the ultrastable laser at Nanshan & 773604 \\
		$f_{CW,NS}$ & Frequency of the ultrastable laser at Nanshan & 193.4THz \\
		\hline
		$f_{CEO,GY}$ & Carrier-envelope offset frequency of the comb at Gaoyazi & -35MHz \\
		$f_{Beat,GY}$ & The beat frequency of the comb at Gaoyazi & 35MHz \\
		$f_{r,GY}$ & The repetition rate of the comb at Gaoyazi & ~250M+2.5kHz \\
		$N_{GY}$ & Index of the comb's tooth locking to the ultrastable laser at Gaoyazi & 773596 \\
		$f_{CW,GY}$ & Frequency of the ultrastable laser at Gaoyazi & 193.4THz \\
		\hline
	\end{tabular}
	\caption{Values of frequency combs and ultrastable lasers at each terminal.}
\end{table}

\begin{figure}[h]
	\centering
	\includegraphics[scale=1]{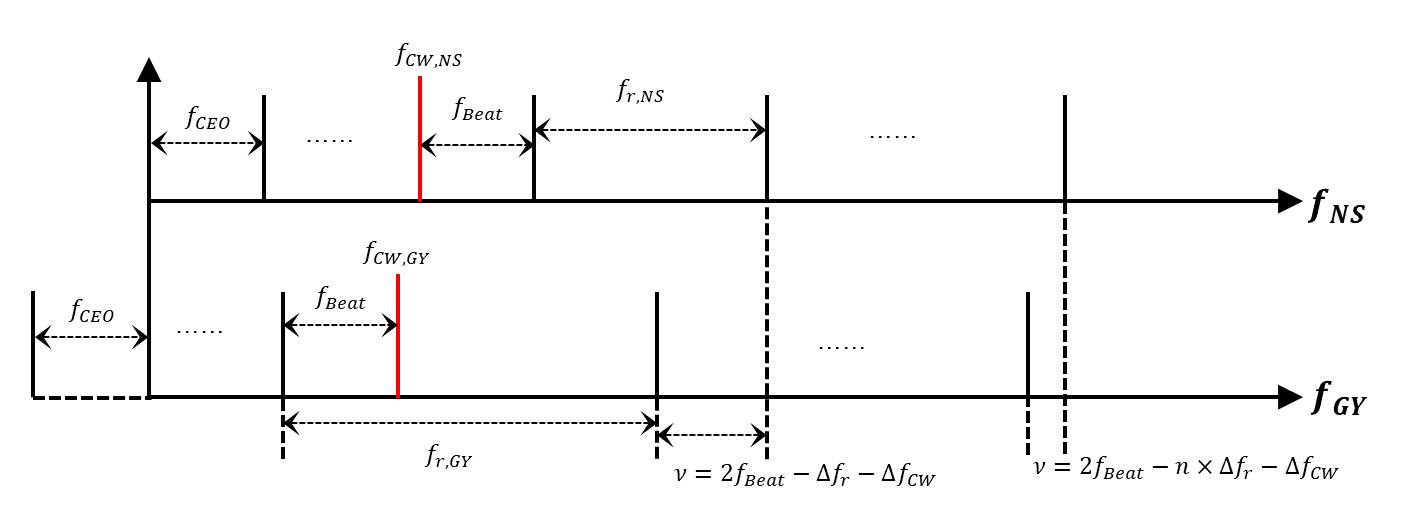}
	\caption{Locking method for optical frequency combs at each terminal.}
\end{figure}

\textbf{Frequency difference calculation and Phase compensation.} 
The electric field of the two combs at terminal A and B can be written as follows:
\begin{equation}
	{{E}_{A}}(t)=\exp (i2\pi {{\omega }_{A}}t)\sum\limits_{n}{{{E}_{A,n}}\exp [i2\pi (n{{f}_{r}}t+{{\varphi }_{A,n}})]}
\end{equation}
\begin{equation}
	{{E}_{B}}(t)=\exp (i2\pi {{\omega }_{B}}t)\sum\limits_{n}{{{E}_{B,n}}\exp [i2\pi ((n{{f}_{r}}+\Delta {{f}_{r}})t+{{\varphi }_{B,n}})]}
\end{equation}
Here, ${{\omega }_{A}}$ and ${{\omega }_{B}}$ are the optical frequencies of one pair of most adjacent teeth of two combs. The difference between ${{\omega }_{A}}$ and ${{\omega }_{B}}$ is smaller than ${{f}_{r}}/2$ . After the low-pass filter with bandwidth ${{f}_{r}}/2$, only the beat of most of the adjacent frequencies is maintained. There will be delay ${{\tau }_{0}}$ when comb B is transmitted to comb A. Thus, the measured voltage at terminal A is as follows:
\begin{equation}
	\begin{aligned}
		V\left(t, \tau_0\right) & \propto \operatorname{Im}\left[E_A^*(t) \cdot E_B\left(t-\tau_0\right)\right] \\
		& =\operatorname{Im} \sum_n\left(E_{A, n}^* E_{B, n} E_{a t m, n}\right) \exp \left\{i 2 \pi \left[\left(\omega_B-\omega_A+n \Delta f_r\right) t\right.\right. \\
		& \left.+\left[-n\left(f_r+\Delta f_r\right) \tau_0-\omega_B \tau_0+\varphi_{B, n}-\varphi_{A, n}+\varphi_{a t m, n}\right]\right\}
	\end{aligned}
\end{equation}
Here, ${{E}_{atm,n}}$ is the atmospheric intensity response, and ${{\varphi }_{atm,n}}$ is the atmospheric phase response. We use $F(\nu ,\tau )$ to represent the Fourier transform of the interferogram, and $\angle F(\nu ,\tau )$ represents the spectral phase. Here $\nu$ is the frequency obtained from the FFT of the interferogram. From Eq. (S.11), we get,
\begin{equation}
	\angle F(\nu ,{{\tau }_{0}})=-2\pi {{\tau }_{0}}\frac{{{f}_{r}}+\Delta {{f}_{r}}}{\Delta {{f}_{r}}}[\nu -({{\omega }_{B}}-{{\omega }_{A}})]-2\pi {{\omega }_{B}}{{\tau }_{0}}+{{\varphi }_{B}}(\nu )-{{\varphi }_{A}}(\nu )+{{\varphi }_{C}}(\nu )
\end{equation}
At different time, the delays are different. Here, we use $\tau $ to represent another delay:
\begin{equation}
	\angle F(\nu ,\tau )=-2\pi \tau \frac{{{f}_{r}}+\Delta {{f}_{r}}}{\Delta {{f}_{r}}}[\nu -({{\omega }_{B}}-{{\omega }_{A}})]-2\pi {{\omega }_{B}}\tau +{{\varphi }_{B}}(\nu )-{{\varphi }_{A}}(\nu )+{{\varphi }_{C}}(\nu )
\end{equation}
The difference between $\angle F(\nu ,\tau )$ and $\angle F(\nu ,{{\tau }_{0}})$ are: 
\begin{equation}
	\begin{aligned}
		\Delta \angle F(\nu ,\tau )&=\angle F(\nu ,{{\tau }_{0}})-\angle F(\nu ,\tau ) \\ 
		& =2\pi \frac{{{f}_{r}}+\Delta {{f}_{r}}}{\Delta {{f}_{r}}}[\nu -({{\omega }_{B}}-{{\omega }_{A}})](\tau -{{\tau }_{0}})+2\pi {{\omega }_{B}}(\tau -{{\tau }_{0}}) 
	\end{aligned}
\end{equation}

Make ${{\tau }_{0}}=0$, and the phase is a linear function to $\nu $, and the slope $2\pi \frac{{{f}_{r}}+\Delta {{f}_{r}}}{\Delta {{f}_{r}}}(\tau -{{\tau }_{0}})$ is proportional to the delay $\tau$.

\textbf{Penalized least-squares algorithm.} 
The measured complex spectral signal $s({\nu})$ can be written as below,
\begin{equation}
	s(\nu)=b\times a +n
\end{equation}
Where $s({\nu})$ is the measured complex spectral signal, $\nu$ is the optical frequency, $b$ is the laser spectrum which is relatively smooth,  $a$ is the molecular resonances, and $n$ is the frequency domain noise.
\begin{equation}
	a=\exp (-\sum\limits_{i=1}^{q}{{{\alpha }_{i}}{{c}_{i}}L})
\end{equation}
Where $q$ is the number of samples; $\alpha$ is the complex absorption rate of 1 ppm molecule; $c$ is path-averaged mole fractions; $L$ is the path length.
Ignore noise, take logarithms on both sides of the above, and make $\sigma =\log (s)$, $\beta =\log (b)$, we get,
\begin{equation}
	\sigma =\beta -\sum\limits_{i=1}^{q}{{{\alpha }_{i}}\cdot {{c}_{i}}\cdot L}
\end{equation}
Write in matrix form,
\begin{equation}
	\left[ \begin{matrix}
		I & -{{\alpha }_{1}} & -{{\alpha }_{2}} & \cdots  & -{{\alpha }_{k}}  \\
	\end{matrix} \right]\left[ \begin{matrix}
		\beta   \\
		{{c}_{1}}\cdot L  \\
		{{c}_{2}}\cdot L  \\
		\vdots   \\
		{{c}_{k}}\cdot L  \\
	\end{matrix} \right]=\sigma
\end{equation}
Make $\left[ \begin{matrix}
	I & -{{\alpha }_{1}} & -{{\alpha }_{2}} & \cdots  & -{{\alpha }_{k}}  \\
\end{matrix} \right]=G$,$\left[ \begin{matrix}
	\beta   \\
	{{c}_{1}}\cdot L  \\
	{{c}_{2}}\cdot L  \\
	\vdots   \\
	{{c}_{k}}\cdot L  \\
\end{matrix} \right]=m$, then we get,
\begin{equation}
	G\cdot m=\sigma
\end{equation}
The above equation is an underdetermined equation with infinitely many solutions. It is necessary to add regularization constraints to determine the unique solution. 
The fidelity of $\sigma$ against $G\cdot m$ can be expressed by the population variance between them: 
\begin{equation}
	F=\sum\limits_{i=1}^{r}{{{({{\sigma }_{i}}-{{(Gm)}_{i}})}^{2}}}={{\left\| \sigma -G\cdot m \right\|}^{2}}
\end{equation}
The roughness of fitted data $\beta$ can be expressed by the sum of squares of its difference:
\begin{equation}
	R=\sum\limits_{i=2}^{r}{{{({{\beta }_{i}}-{{\beta }_{i-1}})}^{2}}}={{\left\| \bm{D}\cdot \beta  \right\|}^{2}}
\end{equation}
Where $\bm{D}$ is the difference matrix. The objective function $Q$ is the sum of the fidelity $F$ and the roughness $R$ with different weights,
\begin{equation}
	Q=F+\lambda \cdot R={{\left\| \sigma -G\cdot m \right\|}^{2}}+\lambda \cdot {{\left\| \bm{D}\cdot \beta  \right\|}^{2}}
\end{equation}
Where $\lambda$ is weighting factor. Let ${{\bm{D}}_{2}}=\left[ \begin{matrix}
	\bm{D} & \bm{0}
\end{matrix} \right]$, we get:
\begin{equation}
	Q={{\left\| \sigma -G\cdot m \right\|}^{2}}+\lambda \cdot {{\left\| {{\bm{D}}_{2}}\cdot m \right\|}^{2}}
\end{equation}
The inversion of concentration and substrate can be written as an optimization problem with constraints. The optimization function is a convex function with a minimum value. At the minimum, the derivative is 0. We get,
\begin{equation}
	\frac{\partial Q}{\partial m}=0
\end{equation}
\begin{equation}
	m={{({{G}^{T}}G+\lambda \cdot \bm{D}_{2}^{T}{{\bm{D}}_{2}})}^{-1}}\cdot {{G}^{T}}\cdot \sigma \text{=}\bm{H}\cdot \sigma
\end{equation}
Where $\bm{H=}{{({{G}^{T}}G+\lambda \cdot \bm{D}_{2}^{T}{{\bm{D}}_{2}})}^{-1}}\cdot {{G}^{T}}$. The difference order is usually set to 2\cite{zhangBaselineCorrectionUsing2010a}, and the weighting factor depends on the smoothness of the substrate. The smoother the base, the greater the weight factor.

\textbf{Deviation analysis of reflected stray light.} The output power is so high that nonlinear effect and scattering are inevitably introduced. Suppose there exists an interferogram that is $\gamma $ times as strong as the original with a delay $\Gamma $, the time domain signal becomes:
\begin{equation}
	{{S}^{'}}(t)=S(t)+\gamma S(t-\Gamma )
\end{equation}
Then the measured complex spectral signal becomes:
\begin{equation}
	{{s}^{'}}(\nu )=(1+\gamma {{\text{e}}^{i2\pi \Gamma \nu }})s(\nu )
\end{equation}
The logarithmic spectrum is:
\begin{equation}
	{{\sigma }^{'}}(\nu )\approx \sigma (\nu )+\log (1+\gamma {{\text{e}}^{i2\pi \Gamma \nu }})
\end{equation}
When $\gamma \ll 1$, $\log (1+\gamma {{\text{e}}^{i2\pi \Gamma \nu }})\approx \gamma {{\text{e}}^{i2\pi \Gamma \nu }}$,so:
\begin{equation}
	{{\sigma }^{'}}(\nu )=\sigma (\nu )+\gamma {{\text{e}}^{i2\pi \Gamma \nu }}
\end{equation}
Therefore, a cosine modulation$\gamma \cos (\pi \Gamma \nu )$ is introduced in the intensity spectrum and a sinusoidal modulation$\gamma \sin (\pi \Gamma \nu )$ is introduced in the phase spectrum.

\end{document}